\documentclass[12pt]{article}
\newcommand{\beq}{\begin{equation}}
\newcommand{\eeq}{\end{equation}}
\begin{document}

\begin{center}

{\bf Two Types of the Effective Mass Divergence and the 
Gr\"uneisen Ratio in Heavy Fermion Metals}\bigskip 

{M.Ya. Amusia$^{a,b}$, A.Z. Msezane$^{c}$, 
and V.R. Shaginyan$^{c,d}$
\footnote{E--mail: vrshag@thd.pnpi.spb.ru}}\\
\bigskip
{\it $^{a\,}$The Racah Institute of Physics,\\
the Hebrew University, Jerusalem 91904, Israel;\\
$^{b\,}$A.F. Ioffe Physical-Technical Institute, 194021 St.\\
Petersburg, Russia;\\
$^{c}$CTSPS, Clark Atlanta University, Atlanta,\\
Georgia 30314, USA;\\
$^{d\,}$Petersburg Nuclear Physics Institute,\\
Gatchina, 188300, Russia}

\end{center}

\begin{abstract}
The behavior of the specific heat $c_p$, effective mass $M^*$, and the
thermal expansion coefficient $\alpha$ of a Fermi system located near the
fermion condensation quantum phase transition (FCQPT) is considered. We
observe the first type behavior if the system is close to FCQPT: the
specific heat $c_p\propto \sqrt{T}$, $M^*\propto1/\sqrt{T}$, 
while the thermal expansion coefficient $\alpha\propto \sqrt{T}$. 
Thus, the Gr\"uneisen ratio 
${\rm \Gamma}(T)=\alpha/c_p$ does not diverges. 
At the transition region, where the system passes over from 
the non-Fermi liquid 
to the Landau Fermi liquid, the ratio 
diverges as ${\rm \Gamma}(T)\propto 1/\sqrt{T}$. When the system becomes
the Landau Fermi liquid, ${\rm \Gamma}(T,r)\propto 1/r$, with $r$ being a
distance from the quantum critical point. 
Provided the system has undergone FCQPT, the second type takes place: the
specific heat behaves as $c_p\propto \sqrt{T}$, $M^*\propto 1/T$, and 
$\alpha=a+bT$ with $a,b$ being constants. 
Again, the Gr\"uneisen ratio diverges
as ${\rm \Gamma}(T)\propto 1/\sqrt{T}$. 
\end{abstract}



\newpage

The nature of the strongly correlated state and the non-Fermi liquid (NFL)
behavior observed in the high-temperature superconductors, heavy-fermion
(HF) metals, two-dimensional (2D) $^3$He and electron systems etc., are
still  hotly debated. It is widely believed that the observed
behavior is determined by quantum phase transitions which occur at quantum
critical points. The proximity of a system to quantum critical points
creates its NFL behavior. The system can be driven to quantum critical
points (QCPs) by tuning some control parameters 
other than temperature, such as 
pressure, magnetic field, or  doping. The NFL behavior observed in many
systems suggests that in principle there can exist many CQPs. The situation
becomes more misty and obscure, if one takes into account that when a system
is close to QCP we are dealing with the strong coupling limit where an
absolutely reliable answer cannot be given, based on purely first
principle theoretical grounds. Therefore, the only way to verify what type of QCP occurs
is to consider experimental facts which describe the behavior of a system.
Only recently, experimental facts have appeared which deliver experimental
grounds to understand the nature of quantum phase transition producing the
NFL behavior of HF metals.

Recent measurements for the HF metal CeNi$_2$Ge$_2$ have shown that the
thermal expansion coefficient $\alpha(T)$ exhibits the NFL behavior, $
\alpha(T)=a\sqrt{T}+bT$, the specific heat $c_p(T)/T=\gamma_0-c\sqrt{T}
+d/T^3$, and that the Gr\"uneisen ratio ${\rm \Gamma}(T)=\alpha(T)/c_p(T)$
diverges as ${\rm \Gamma}(T)\propto 1/\sqrt{T}$ \cite{geg}. Here, $
a,b,\gamma_0,c,d$ are constants, and the low-temperature upturn in the
specific heat related to the $d/T^3$ term can be ascribed to the
high-temperature tail of a Shotky anomaly, whose influence on the measured
Gr\"uneisen ratio is small \cite{geg}. While measurements on the HF metal
YbRh$_2($Si$_{0.95}$Ge$_{0.05}$)$_2$ have revealed the different NFL
behavior: at $T<1$ K, the thermal expansion coefficient 
$\alpha(T)=a_1+a_0T$, 
and the Gr\"uneisen ratio diverges as ${\rm \Gamma}(T)\propto 1/T^{z}$,
where the exponent $z$ is fractional, $z=0.7\pm 0.1$ \cite{geg}.

A general consideration based on the assumption of scaling and a model study
on the spin-density-wave \cite{zhu} has produced an explanation
for the NFL behavior of the thermal expansion coefficient in the case of 
CeNi$_2$Ge$_2$ but failed to explain the NFL behavior observed in 
YbRh$_2($Si$_{0.95}$Ge$_{0.05}$)$_2$. 
As a result, a challenge still remains how to
explain within one conceptual framework the NFL behavior observed in these HF
metals, both of which are considered to be close to QCPs.

In this Letter, we show that within the framework of the fermion
condensation quantum phase transition (FCQPT) \cite{ms,mms}, which leads to the
formation of fermion condensate (FC) \cite{ks}, it is possible to understand
the NFL behavior observed in different strongly 
and highly correlated Fermi systems
such as high-$T_c$ superconductors \cite{ms,mms,ks,ams}, heavy-fermion
metals \cite{mms,shag,shag1}, and 2D systems \cite{mms,shag1,shag2}. We
consider the specific heat $c_p$, effective mass $M^*$, and the thermal
expansion coefficient $\alpha$ of a Fermi system near FCQPT and
show that at sufficiently high temperatures, 
the specific heat $c_p\propto \sqrt{T}$, $M^*\propto1/\sqrt{T}$, and 
the thermal expansion coefficient $\alpha(T)\propto \sqrt{T}$. Thus, the
Gr\"uneisen ratio ${\rm \Gamma}(T)=\alpha/c_p$ does not diverge. 
At the transition temperatures, when the system comes from the 
NFL behavior to the Landau Fermi liquid (LFL) one, the ratio diverges as 
${\rm \Gamma}(T)\propto 1/\sqrt{T}$. 
As the system possesses the LFL behavior at low temperatures, 
${\rm \Gamma}(T,r)\propto 1/r$, here $r$ is a distance from FCQPT.  
If the system has undergone FCQPT and FC is formed,
the specific heat behaves as $c_p\propto \sqrt{T}$, and $\alpha=a+bT$ with $
a,b$ are constants. Again, the Gr\"uneisen ratio diverges as ${\rm \Gamma}
(T)\propto 1/\sqrt{T}$. Thus, the second type of behavior can be observed 
in a Fermi system with FC, which nonetheless demonstrates 
the main features of the critical behavior, 
as if the system were located near QCP. Our results are in good agreement with
measurements on single crystals of 
CeNi$_2$Ge$_2$ and YbRh$_2($Si$_{0.95}$Ge$_{0.05}$)$_2$ \cite{geg}.
We predict that if the system is driven back to LFL by the application of a magnetic
field $B$, then in the case of YbRh$_2($Si$_{0.95}$Ge$_{0.05}$)$_2$ 
the thermal expansion coefficient 
$\alpha(B)\propto 1/\sqrt{B}$, while in the case of CeNi$_2$Ge$_2$  the
coefficient behaves as $\alpha(B)\propto B^{-2/3}$. 

Let us start by considering the key points of the FC theory. We concentrate
on the 3D case because heavy fermion metals are 3D structures 
\cite{kad}. On
the other hand, there is no a qualitative difference between 3D and 2D
cases in the theory of FC, and our results are applicable to both
2D and 3D cases \cite{shag,shag1}. 
FC represents a new solution of the Fermi
liquid theory equations \cite{lan} for the quasiparticle distribution
function $n({\bf p},T)$ 
\begin{equation}
\frac{\delta (F-\mu N)}{\delta n({\bf p},T)}
=\varepsilon ({\bf p},T)-\mu (T)-T\ln 
\frac{1-n({\bf p},T)}{n({\bf p},T)}=0,
\end{equation}
which depends on the momentum ${\bf p}$ and temperature $T$. 
Here $F=E-TS$ is the
free energy, and $\mu $ is the chemical potential, 
while $\varepsilon ({\bf p},T),$
\begin{equation}
\varepsilon ({\bf p},T)=\frac{\delta E[n(p)]}{\delta n({\bf p},T)},
\end{equation}
is the quasiparticle energy. This energy is a functional of $n({\bf p},T)$
just like the total energy $E[n(p)]$, entropy $S[n(p)]$ and the other
thermodynamic functions. The entropy $S[n(p)]$ is given by the familiar
expression 
\begin{equation}
S[n(p)]=-2\int \left[ n({\bf p},T)\ln n({\bf p},T)+(1-n({\bf p},T))\ln (1-n(
{\bf p},T))\right] \frac{d{\bf p}}{(2\pi )^3},
\end{equation}
which stems from purely combinatorial considerations. Eq. (1) is usually
presented as the Fermi-Dirac distribution  
\begin{equation}
n({\bf p},T)=\left\{ 1+\exp \left[ \frac{(\varepsilon ({\bf p},T)-\mu )}T
\right] \right\} ^{-1}.
\end{equation}
At $T\to 0$, one gets from Eqs. (1) and (3) the standard solution 
$n_F({\bf p},T\to 0)\to \theta (p_F-p)$, 
with $\varepsilon (p\simeq p_F)-\mu
=p_F(p-p_F)/M_L^{*}$, where $p_F$ is the Fermi momentum, 
$\theta (p_F-p)$ is the step function,  
and $M_L^{*}$ is the Landau effective mass \cite{lan}  
\begin{equation}
\frac 1{M_L^{*}}=\frac 1p\frac{d\varepsilon (p,T\to0)}{dp}|_{p=p_F}.
\end{equation}
It is implied that in the case of LFL $M_L^{*}$ is
positive and finite at the Fermi momentum $p_F$. As a result, the 
$T$-dependent corrections to $M_L^{*}$, to the quasiparticle energy 
$\varepsilon (p)$, and other quantities, start with $T^2$-terms. But this
solution is not the only one possible. There exist solutions of Eq. (1)
associated with the so-called fermion condensation \cite{ks}. Being
continuous and satisfying the inequality $0<n_0({\bf p})<1$ within some 
region in $p$, such solutions $n_0({\bf p})$ 
admit a finite limit for the logarithm in Eq.
(1) at $T\rightarrow 0$ yielding  \cite{ks} 
\begin{equation}
\varepsilon ({\bf p})-\mu \ 
=\ 0,\quad \mbox{if}\quad 0<n_0({\bf p})<1;\,p_i\leq p\leq p_f\ .
\end{equation}
At $T\to 0$, Eq. (6) defines a new state of electron liquid with FC \cite
{ks,vol} which is characterized by a flat part of the spectrum in the 
$(p_f-p_i)$ region and can strongly 
influence measurable quantities up to temperatures 
$T\ll T_f$ \cite{ms,ks}. Note, that a formation of the flat part of the spectrum
has been recently confirmed \cite{dzy,irk}, and the momenta 
$p_i$ and $p_f$ have to satisfy $p_i<p_F<p_f$.
At $T\to 0$, Eq. (6) defines a particular state of a Fermi liquid
with FC, for which the order parameter $\kappa ({\bf p})=\sqrt{(1-n_0({\bf p}
))n_0({\bf p})}$ has finite values in the $(p_f-p_i)$ region, whereas the
superconducting gap $\Delta _1\to 0$ in this region. Such a state can be
considered as superconducting, with an infinitely small value of 
$\Delta _1$, so that the entropy $S(T=0)$ of this state 
is equal to zero. It is obvious that
this state being driven by the quantum phase transition disappears at $T>0$ 
\cite{ms}. When the density $x$ tends to QCP located at $x_{FC}$, $x\to x_{FC}$, 
the Landau amplitude $F_L(p=p_F,p_1=p_F,x)$ as
a function of the density $x$ becomes smaller, and the flat part vanishes.
Then, at $T\to 0$, Eq. (6) has the only trivial solution 
$\varepsilon (p=p_F)=\mu $, and the 
quasiparticle occupation numbers are given by the step function, 
$n({\bf p})=\theta (p_F-p)$ \cite{ks}. 

It follows from Eq. (6) that at 
$T_f\gg T>0$, the system becomes divided into two quasiparticle subsystems:
the first subsystem is occupied by normal quasiparticles with 
the finite effective mass 
$M_L^{*}$ independent of $T$ at momenta $p<p_i$, and the second subsystem in
the $(p_f-p_i)$ range is characterized by the quasiparticles with the
effective mass $M_{FC}^{*}(T)$ \cite{ms,khod1} 
\begin{equation}
M_{FC}^{*}\simeq p_F\frac{p_f-p_i}{4T}.
\end{equation}
There is the energy scale $E_0$ separating the slow dispersing low energy
part, related to the effective mass $M_{FC}^{*}$, from the faster dispersing
relatively high energy part, defined by the effective mass $M_L^{*}$. It
follows from Eq. (7) that $E_0$ is of the form \cite{ms}  
\begin{equation}
E_0\simeq 4T.
\end{equation}
The described system can be viewed as the strongly correlated system 
and has the second type of the behavior (the first type will be considered below). 
At $T\to 0$, the system demonstrates the typical
critical behavior as if the system were close to a quantum critical point:
the effective mass diverges as $M^{*}\propto 1/T$, while the application of
magnetic field $B$ restores the LFL behavior with the effective mass 
$M^{*}\propto 1/\sqrt{B}$ \cite{shag}. Note that such a behavior is in
agreement with experimental facts \cite{geg1}.

The existence of the two quasiparticle subsystems can be illuminated by
calculating the thermal expansion coefficient $\alpha (T)$ \cite{khod1},
which is given by \cite{lanl1} 
\begin{equation}
\alpha (T)=\frac 13\left( \frac{\partial (\log V)}{\partial T}\right) _P
=-\frac x{3K}\left( \frac{\partial (S/x)}{\partial x}\right) _T.
\end{equation}
Here, $P$ is the pressure and $V$ is the volume. The compressibility $K$ is
not expected to be singular at FCQPT and in systems with FC, because FC is
attached to the Fermi level, and it moves along as $\mu (x)$ changes, while
the compressibility $K=d\mu /d(Vx)$ is approximately constant \cite{noz}.
Inserting Eq. (3) into Eq. (9), we find that 
\begin{equation}
\alpha (T)=\frac{S(T)}{3Kx}-\frac{1}{3K}\int \frac{\partial n({\bf p},T)}
{\partial x}\ln \frac{1-n({\bf p,T})}{n({\bf p},T)}\frac{d{\bf p}}{(2\pi )^3}.
\end{equation}
Because the region $(p_f-p_i)$ is occupied by FC,  
we have from Eq. (1) that the ratio 
$\beta =(\varepsilon({\bf p})-\mu )/T\simeq \ln (1-n_0({\bf p}))/n({\bf p})$
does not depend on the temperature. We expect corrections to be of the order 
$T$ at $T_f\gg T$, so that $\beta \simeq const+T$. Thus, the
right hand side of Eq. (10) can be estimated as $\propto a+bT$, the term $a$
is determined by the FC contribution, and $bT$ is given by normal
quasiparticles with the effective mass $M_L^{*}$. 
Therefore, in the case of systems with FC,
the thermal expansion coefficient $\alpha _{FC}=a+bT$, while in the case of
LFL, one obtains from Eq. (10) the standard expression $\alpha (T)\sim
M_L^{*}T/p_F^2K.$ On the other hand, for the two quasiparticle subsystems
with $M_{FC}\propto 1/T$, see Eq. (7), Eq. (10) translates into 
\begin{equation}
\alpha _{FC}(T)\sim \frac{M_{FC}^{*}T}{p_F^2K}+\frac{M_L^{*}T}{p_F^2K}
\propto a+bT.
\end{equation}
We again find that $\alpha _{FC}(T)=a+bT$. This behavior of the thermal
expansion coefficient predicted in Ref. \cite{khod1} is in good agreement
with experimental measurements on YbRh$_2($Si$_{0.95}$Ge$_{0.05}$)$_2$ \cite
{geg}. 

The application of magnetic fields to YbRh$_2($Si$_{0.95}$Ge$_{0.05}$)$_2$ 
restores the LFL behavior, and the effective mass  scales as
$M^*\propto 1/\sqrt{B}$ at sufficiently low temperatures $T<T^*(B)$,
where $T^*(B)\propto \sqrt{B}$ is the temperature above which 
the system comes back to the NFL behavior \cite{shag}. 
As a result, we have from Eq. (10) 
that the thermal expansion coefficient depends on the magnetic field
as $\alpha(B)\propto 1/\sqrt{B}$.

According to the Landau theory, the specific heat per unit volume is 
\begin{equation}
c(T)=2\int (\varepsilon ({\bf p})-\mu )\frac{\partial n({\bf p})}{\partial T}
\frac{d{\bf p}}{(2\pi )^3}.
\end{equation}
At low temperatures, as in the common case, we can ignore 
the difference $\delta c$ between 
$c_v$ and $c_p$ in systems with FC because 
$(c_v(T)-c_p(T))\propto T\alpha^2(T)$ \cite{lanl1}, 
and as it follows from the above, 
$\delta c \propto T$ \cite{khod1}. 
On the other hand, this difference is large when compared to the common case
for $\delta c\propto T^3$ \cite{lanl1}.
As we have seen above, the ratio $\beta $ is approximately 
constant in the region $(p_f-p_i)$. Using this fact and 
differentiating both sides of Eq. (4) with respect to $T$, we can
check that $\partial n({\bf p})/\partial T\propto T$. As a result, it is seen
from Eq. (12) that the FC region does not contribute to the 
singular behavior of the specific heat.
The contribution coming from the normal quasiparticles with $M_L^{*}$ is of
the standard form being proportional to $T$. There is a specific contribution
related to the spectrum $\varepsilon ({\bf p})$ which insures the connection
between the dispersionless region $(p_f-p_i)$ occupied by FC and the normal
quasiparticles located at $p<p_i$ and at $p>p_f$, and therefore it is of the
form $\varepsilon ({\bf p})\sim (p-p_f)^2\sim (p_i-p)^2$. 
Such a form of the spectrum can
be verified in exactly solvable models for systems with FC \cite{ks}. 
One can check, that the
contribution of this spectrum to $c_p\propto \sqrt{T}$,
and this consequence coincides with one obtained in exactly solvable 
models \cite{ks}. As a
result, we have $c_p(T)\propto a_1\sqrt{T}+b_1T$, with $a_1,b_1$ being some
constants. It is worth noting that when calculating quantities like $c_p$
containing the derivative $\partial n({\bf p})/\partial T$, it is incorrect
to insert Eq. (7) into the standard formula for the specific heat, 
$c(T)\propto M^{*}T$. 
Now we can calculate that the Gr\"uneisen ratio diverges as 
${\rm \Gamma }(T)\propto 1/\sqrt{T}$. 
The obtained Gr\"uneisen exponent $z=1/2$ is
close to the experimental one $z=0.7\pm 0.1$ \cite{geg}. We conclude that in
contrast to the common belief, the heavy-fermion metal 
YbRh$_2($Si$_{0.95}$Ge$_{0.05}$)$_2$ 
need not be located very near CQP, and there is no
need to introduce a new type of QCP to explain its properties.

Consider the case when the system is close to FQCPT and demonstrates the first
type of the behavior. FCQPT can be induced by tuning the density $x$. When
the system approaches FCQPT from disordered phase, FCQPT manifests itself in
the divergence of the quasiparticle effective mass $M^*$ as the density 
tends to the critical density 
$x_{FC}$, or the distance $r\sim |x-x_{FC}|\to 0$ \cite{shag1,khod} 
\begin{equation}
M^*\propto \frac{1}{|x-x_{FC}|}\propto \frac{1}{r}.
\end{equation}
Note, that Eq. (13) is valid in both cases of 2D and 3D systems 
\cite{shag1}. Since the effective mass 
$M^*$ is finite, the system exhibits the LFL
behavior at low temperatures $T\sim T^*(x)\propto|x-x_{FC}|^2$ \cite{shag1}. 
At sufficiently high temperatures, the system
possesses the NFL behavior. At $|x-x_{FC}|/x_{FC}\ll 1,$
this behavior can be viewed as the highly correlated one, because the
effective mass strongly depends on the control parameters such as the
density, temperature and magnetic fields \cite{shag1}. In contrast to the
strongly correlated system, the highly correlated system does not have the
energy scale given by Eq. (8), Eq. (7) is not also valid, and at $T<T^*(x)$
it behaves as LFL \cite{shag1}. Let us calculate the Gr\"uneisen ratio at 
$T\sim T^*(x)$. The entropy $S$ is given by the LFL formula, $S\propto M^*T$.
Inserting this into Eq. (9) and taking into account Eq. (13), we
obtain $\alpha(T)\propto T/|x-x_{FC}|^2$ while $c_p\propto T/|x-x_{FC}|$. 
As a result, the Gr\"uneisen ratio diverges as 
\begin{equation}
{\rm \Gamma}(T,r)\propto \frac{1}{|x-x_{FC}|}\propto \frac{1}{r}\propto M^*.
\end{equation}
At the transition temperatures, the system passes from
the NFL behavior to the LFL one, and the effective mass depends on both
$T$ and $x$. In this case, we expect  Eq. (14) to hold true because
the effective mass has to be continuous  over the transition region.

Now consider the Gr\"uneisen ratio at $T\gg T^*(x)$ when the effective mass
depends predominantly on the temperature. 
Landau equation relating the mass $M$ of an
electron to the effective mass of the quasiparticles is of the form 
\cite{lan} 
\begin{equation}
\frac{{\bf p}}{M^*}=\frac{{\bf p}}{M}+ \int F_L({\bf p},{\bf p}_1,x)
\nabla_{{\bf p}_1}n({\bf p_1}) \frac{d{\bf p}_1}{(2\pi)^3}.
\end{equation}
Applying Eq. (15) at $T<T^*(x)$, we obtain the common result 
\begin{equation}
M^*=\frac{M}{1-N_0F^1_L(x)/3}.
\end{equation}
Here $N_0$ is the density of states of a free Fermi gas and 
$F^1_L(x)$ is the $p$-wave component of the Landau interaction. 
At $x\to x_{FC}$, the
denominator in Eq. (16) tends to zero and one obtains Eq. (13). The
temperature smoothing of the step function $\theta(p_F-p)$ at 
$p_F=(x/3\pi^2)^{1/3}$ induces the variation of the 
Fermi momentum $\Delta
p_F\sim TM^*/p_F$. We assume that the amplitude $F_L$ 
has a short range $q_0\ll p_F$ of interaction in the momentum space,
which is a common condition leading to the existence of FC 
and nearly-localized Fermi liquids \cite{ks,pfi}. 
If the radius is that $q_0\sim\Delta p_F\sim T_0M^*/p_F$, 
corrections to the effective
mass are proportional to $T$ at $T\sim T_0$. Here 
$T_0\propto |x-x_{FC}|$ is a characteristic temperature 
at which the system's behavior is the NFL one. 
On the other hand, since $T^*(x)\ll T_0$,
we have $q_0\gg T^*(x)M^*/p_F$, and the system behaves as LFL  at
$T\sim T^*(x)$, that is corrections to the effective mass start with 
$T^2$ terms. We can also conclude that the transition region is rather
large compared with $T^*(x)$, being proportional to $T_0$.

In the case of $T\sim T_0$, we again can use Eq. (16) with 
$F^1_L(p_F+\Delta p_F)\sim F^1_L(p_F)+A\Delta p_F$, 
where $A\propto dF^1_L(x)/dx$.
Substituting this expansion of $F^1_L(p_F+\Delta p_F)$ into Eq. (16) we find
that \begin{equation}
M^*\sim\frac{M}{A\Delta p_F}\propto \frac{M}{M^*T}.
\end{equation}
In deriving Eq. (17) it is implied that the system is close to FCQP so that 
$(1-N_0F^1_L(p_F))\ll N_0A\Delta p_F$. 
We can say that at $T\sim T_0$, $\Delta
p_F$ induced by $T$ becomes larger then the distance $r$ from FCQP, 
and $\Delta p_F>|p_F^{FC}-p_F|$, where $p_F^{FC}$ corresponds to $x_{FC}$. 
As a result, Eq. (17) becomes 
\begin{equation}
M^*\propto \frac{1}{\sqrt{T}}.
\end{equation}
Then from Eqs. (10) and (18), we obtain that at $T\sim T_0$, the main
contribution to the thermal expansion coefficient 
$\alpha(T)\propto a_1\sqrt{T}+b_1T$ in accordance with experimental facts \cite{geg}. 
Here the $b_1$ term comes from the contribution of normal quasiparticles.

By applying magnetic fields the system in consideration 
can be driven back to LFL with
the effective mass $M^*\propto B^{-2/3}$ at temperatures $T<T^*(B)$,
where $T^*(B)\propto B^{4/3}$ \cite{shag1}. Thus, in the case of
CeNi$_2$Ge$_2$, it follows from Eq. (10) that the thermal expansion coefficient 
depends on the magnetic field as $\alpha(B)\propto B^{-2/3}$. 
We note that the magnetic field should not be too high, otherwise
the distance $r$ becomes large, and the system  cannot be considered to 
be close to the quantum critical point. A more detailed analysis of this issue
will be published elsewhere.

When considering the specific heat, we again ignore the difference 
$(c_p(T)-c_v(T))\propto T\alpha^2(T)\propto T^2$.
Note that this difference is large in comparison to the common case. 
It follows from Eq. (12) that the specific heat 
$c_p(T)\simeq c_v(T)\propto \sqrt{T}$. This result implies that 
the Gr\"uneisen ratio does not diverge at $T\sim T_0$. 
However, at the transition region where the effective mass depends on both
$T$ and $x$, it is seen from Eq. (14) that the ratio diverges,  
${\rm \Gamma}(T)\propto 1/\sqrt{T}$, and finally at $T\sim T^*(x)$,
we have ${\rm \Gamma}(T,r)\propto 1/|x-x_{FC}|\propto 1/r$.  

It follows from measurements on CeNi$_2$Ge$_2$ that 
${\rm \Gamma}(T)\propto 1/\sqrt{T}$, while 
the behavior of the specific heat 
$c_p(T)\propto \gamma_0T-cT^{3/2}$ \cite{geg} suggests
that the system in question enters the transition region.  Both these
observations are consistent with our consideration. Therefore,
we expect that at decreasing temperatures the ratio will be saturated. 

In conclusion, we have shown that our simple model based on FCQPT explains the
critical behavior of the thermal expansion coefficient, specific heat and the
Gr\"uneisen ratio observed in different heavy fermion metals. In the case of 
CeNi$_2$Ge$_2$, the behavior of 
these quantities can be explained by its closeness to
FCQPT, while the electronic system behaves as a highly correlated liquid.
In the case of YbRh$_2($Si$_{0.95}$Ge$_{0.05}$)$_2$ the behavior of these
quantities is different, and can be explained by the presence of FC in
this metal, that is the electronic system has undergone FCQPT and behaves
as a strongly correlated liquid. 
We also predict that if the system is driven back to LFL by the application of a magnetic
field $B$, then in the case of a strongly correlated liquid
the thermal expansion coefficient 
$\alpha(B)\propto 1/\sqrt{B}$, while in the case of a highly correlated liquid   
the coefficient behaves as $\alpha(B)\propto B^{-2/3}$. 

The visit of VRS to Clark Atlanta University has been supported by NSF
through a grant to CTSPS. MYaA is grateful to the S. A. Shonbrunn Research
endorment Fund for support of his research. AZM is supported by US DOE,
Division of Chemical Sciences, Office of Basic Energy Sciences, Office of
Energy Research. This work was supported in part by the Russian Foundation
for Basic Research, No 04-02-16136.

\end{document}